# Human-Machine Networks: Towards a Typology and Profiling Framework

Aslak Wegner Eide[1], J. Brian Pickering[2], Taha Yasseri[3], George Bravos[4,5], Asbjørn Følstad[1], Vegard Engen[2], Milena Tsvetkova[3], Eric T. Meyer[3], Paul Walland[2], Marika Lüders[1]

[1]SINTEF, Oslo, Norway
`{aslak.eide,asbjorn.folstad,marika.lüders}@sintef.no`
[2]IT Innovation Center, University of Southampton, Southampton, UK
`{jbp,ve,pww}@it-innovation.soton.ac.uk`
[3]Oxford Internet Institute, University of Oxford, Oxford, UK
`{taha.yasseri,milena.tsvetkova,eric.meyer}@oii.ox.ac.uk`
[4]Athens Technology Center, Athens, Greece
[5]Hellenic American University, Athens, Greece
`g.bravos@atc.gr`

**Abstract.** In this paper we outline an initial typology and framework for the purpose of profiling *human-machine networks*, that is, collective structures where humans and machines interact to produce synergistic effects. Profiling a human-machine network along the dimensions of the typology is intended to facilitate access to relevant design knowledge and experience. In this way the profiling of an envisioned or existing human-machine network will both facilitate relevant design discussions and, more importantly, serve to identify the network type. We present experiences and results from two case trials: a crisis management system and a peer-to-peer reselling network. Based on the lessons learnt from the case trials we suggest potential benefits and challenges, and point out needed future work.

## 1 Introduction

The world we live in is suffused with interconnected information and communication technology (ICT) components that have become a ubiquitous part of virtually every aspect of our daily lives. At work and in our private lives, when we socialize, create, collaborate or play, we often do so in networks of both humans and machines (e.g. in online social networks, online retail platforms, collaboration platforms, crowdsourcing engines, decision support systems, and massive multiplayer games). Designing and developing for such human-machine networks (HMNs) poses immense challenges.

As technologies and services are integrated into interacting networks of humans and machines rather than being taken up by individual users, classical approaches to human-centred design (HCD) may no longer provide a sufficient degree of design support. This is for example borne out by the challenges involved in establishing sustainable social networks [1], collaborative systems for knowledge workers [2], and citizen-government collaboration systems [3].

To improve the general understanding of HMNs and strengthen our capability to design for such networks, we are developing an HMN typology and associated profiling framework that can be used by designers and developers during the HCD development cycle. A typology is an approach to classification, where the basic concept is detailed according to its salient dimensions [4]. The profiling framework is intended to support access to design knowledge and experience on the basis of the typology.

This paper describes the initial versions of the typology and profiling framework, and illustrates their usefulness through case examples on real-life HMNs. We first present the background on which the typology is based. We then set out the objectives for the typology and profiling framework, before describing their current, initial versions. Finally we present two case trials in which the typology and profiling framework have been applied, before discussing lessons learnt, limitations, and future work.

The typology and associated profiling framework are developed as part of the HUMANE project (http://humane2020.eu). A comprehensive presentation of the typology and framework can be found in the HUMANE technical report *Typology and method v1* [5].

## 2   Background

In human-centred design, as described in the relevant international standard [6], much emphasis is put on the context analysis and requirements phases of development. However, whereas context analysis and requirements in HCD is skewed towards understanding and specifying the required interactions between individual users and machine interfaces, design for human-machine networks needs strengthened support for identifying and modelling the entire network during these phases. Hence, we need design support that allows human-centred designers and future thinkers to benefit from existing knowledge and experience on the level of HMNs.

A number of theoretical concepts have been developed to understand aspects of what we term HMNs. One example is the theory of socio-technical systems, which provides insight into the dual shaping of technology and the social (work) context in which it is implemented, recognizing organizations as complex systems of humans and technology [7]. Another example, actor-network theory, argues that we explicitly need to take into account that any social system is an association of heterogeneous elements such as humans, norms, texts, devices, machines, and technology, thus granting equal weight to humans and non-human (machine) entities in the analysis of the social [8]. A third and newer perspective, the study of social machines, focuses on online systems that combine social participation with machine-based computation, connecting with Berners-Lee's original vision of the Web more as a social creation than a technical one [9]. Though insightful, these theories tend towards a narrow scope, too restrictive to provide a unified framework for understanding HMNs. Instead, selecting high-level constructs of interactive synergies resulting from the behaviours of all actors, we may begin to develop a unified approach.

To establish the background required for developing the HMN typology, we need to define the term and its scope. Based on a systematic literature review across fields of research that attempt to conceptualize networks comprising both humans and machines [10], we defined HMNs as networks in which the behaviours of different actors result in synergistic effects. That is, in human-machine networks, the interaction of human and machine actors allows for objectives to be set and met that would not be feasible without such networked interaction.

The review suggested four analytical layers for studying HMNs: actors, interactions, networks, and behaviours. Actors are the nodes in the HMNs. We distinguish between human actors, who may be represented by individuals, organizational roles, or entire organizations, and machine actors, which may be represented by single devices, as well as by complex back-end systems, as long as they behave in the HMN as a single node. The human and machine actors interact in the HMN. Thus, at the interaction layer, we focus on the (mediated) human-human interactions, human-machine interactions, and machine-machine interactions. The network layer concerns the integration of actors and interaction into larger compounds and aims towards defining types of such sets of actors and interactions. The behavioural layer concerns the emergent qualities of HMNs. Among these are the changing characteristics or roles of actors depending on network context, emergence of new patterns of interaction in the HMN, new applications of the network, and the overall evolution of the network.

As explained further below, the proposed typology was built upon the four layers of actors, interactions, networks, and behaviours.

## 3    Objectives

The main aim of this paper is to present an initial typology and framework towards efficient and accurate profiling of human-machine networks. Profiling a human-machine network is intended to identify the network type, thereby facilitating access to relevant design knowledge and experience where successful HMNs are analysed for the purpose of reusing generic design solutions contributing to their success [11]. Moreover, the profiling's process aims to facilitate relevant design discussion, where the stakeholders may address a richer set of aspects pertaining to the network of humans and machines as part of the context analysis and requirements phases than what is typically done today.

To that end, the typology on which the profiling framework is based should include the key dimensions to support HMN design decisions; that is, the dimensions that ICT-developers typically need to consider in analysis, design, and evaluation. Furthermore, the dimensions should clearly discriminate between different human-machine networks of individual ICT projects. The key objectives of the HMN typology and profiling framework are to:

(1) Help design teams reflect upon system characteristics at the level of networks of humans and machines,
(2) Support creation of a profile of the HMN that works as a documentation of its envisioned network characteristics,
(3) Enable designers to identify relevant successful HMNs,
(4) Support elicitation of relevant design implications.

## 4 HMN Typology and Framework

### 4.1 Developing the Typology

This first version of the HMN typology was developed following the steps outlined in Fig. **1**. Through the initial literature review [10], key constructs were identified and analytical layers established. The literature review also provided insight into relevant dimensions for classification. Then, an initial set of dimensions were suggested and applied to six cases during a workshop. The workshop experiences suggested the initial dimensions be refined into an interim set of dimensions, which was then applied by way of an initial profiling for each of those same cases. Following from the experiences of this initial profiling, a refined set of dimensions were established. This refined set constitutes the first version of the HMN typology.

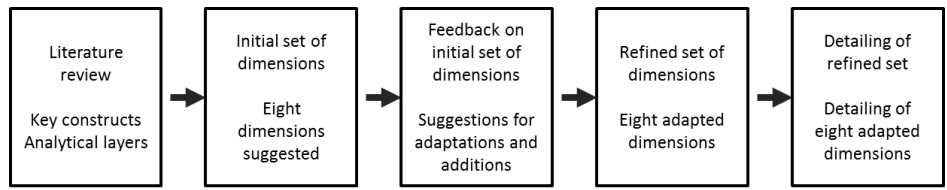

**Fig. 1.** Overview of the process leading to the initial HMN typology and framework

### 4.2 The Typology

We structured the typology based on the four analytical layers (actors, interactions, network, and behaviours) detailed in Section 2. We assigned two dimensions to each layer representing key defining characteristics of HMNs. The typology dimensions are given in Fig. **2**.

The "actors" layer comprises two dimensions – Human agency and Machine agency. Agency is the capacity of the actors in terms of what they can do and achieve in the network. The two dimensions facilitate distinctions between HMNs characterized by varying degrees of automation, artificial intelligence and robot actors, as well as HMN characterized by active collaboration between human and machine agents. We do not imply that machines can exhibit agency on the same level as humans. However, it is practical for the purposes of this typology to refer to machine agency, especially as machines may influence and affect the agency in human actors [12].

At the "interactions" layer, we consider the strength of human to human ties and human to machine interactions as the two dimensions. Tie strength between humans is important for familiarity and trust on the one hand, and social heterogeneity and access to complementary skills and knowledge on the other. While tie strength is much explored, human-machine interaction strength is less studied. However, as machine agency increases, e.g., with the increasing use of social robots, the interaction strength between humans and machines may become increasingly important to HMNs, as is for example seen in some health-care areas [13].

| Layer | Dimension | Description | Scale |
| --- | --- | --- | --- |
| Actors | 1. Human agency | The capacity of the human actors in terms of what they can do and achieve in the network. | low, intermediate, high |
| | 2. Machine agency | The capacity of the machine actors in terms of what they can do in the network, as well as to what extent they enable agency in human actors. | low, intermediate, high |
| Interaction | 3. Tie strength | The tie strength between human nodes in the network. | no ties, latent ties, weak ties, strong ties |
| | 4. H2M interaction strength | The nature and strength of the interaction between humans and machines (H2M) in the network. | independent-optional, independent-necessary, reliant-optional, reliant-necessary |
| Network | 5. Network size | The number of human nodes in the network. | small, medium, large, massive |
| | 6. Geographical space | The geographical extension of the network. | local, regional, global |
| Behaviours | 7. Workflow interdependence | The level of interdependence between actors of in the network. | low, intermediate, high |
| | 8. Network organization | Network organization with implications for predictability and emergence. | bottom-up, intermediate, top-down |

**Fig. 2.** Overview of the typology dimensions

Several dimensions may be considered as critical at the "network" level of HMNs. In the initial version of the typology, we have chosen to include the dimensions of network size and geographical space, due to their importance for the sustainability of HMNs. Growth in terms of size and spread are today seen as key objectives for many HMNs, seen particularly for social networks such as Facebook and Twitter that rely on network effects both for functional and commercial reasons. Furthermore, network size and geographical space have important implications for other dimensions of the network, such as the need for increased machine agency and decreasing of social tie strength with increasing network size.

Finally, at the "behaviours" layer, HMNs are characterized by their workflow interdependence and network organization. Both dimensions concern networks' capacity for emergent change. The former concerns the degree to which the actions of the actors in the network are dependent on and need to be synchronized with the actions of others. The latter concerns the degree of bottom-up vs. top-down organization of the network, where in top-down networks organization is imposed and controlled whereas in a bottom-up network it is self-organising and organic. For example, emergent change may be more prevalent in networks characterized as bottom-up, where initiatives may spread from the grassroots. While efficient spread and refinement of emerging practices may require a certain level of interdependence between the network actors.

### 4.3 The Profiling Framework

The typology and dimensions alone provide limited support for HCD. To facilitate use by others, we have developed a profiling framework as summarized in **Fig. 3**.

| | 1. Define network characteristics | 2. Create joint network profile | 3. Identify similar networks | 4. Extract design principles |
|---|---|---|---|---|
| **Steps** | Describe and scope the characteristics of the network according to the eight dimensions in the typology | Create a joint profile of the network based on the network characteristics defined in step 1. | Identify similar networks based on the network profile generated in step 2. | Extract design principles (design implications, design patterns) from similar networks |
| **Output** | Document describing network characteristics | Network profile | List of similar networks | Collection of design principles |

**Fig. 3.** Overview of profiling framework procedure

The profiling framework comprises four steps, during which various levels of interaction and discussion are possible, and lead to a set of comparative descriptors for a given HMN along with some indication of common HCD design features and issues. Step 1, as shown above, involves an initial estimate of the overall characteristics of the network as defined in regard to the four analytical layers and associated dimensions. Human agency, for example, may be seen as high, but machine agency as low or intermediate; this would relate to a network where most activity is initiated by human actors, with technology components simply responding to their requests. Once some indication of values associated with dimensions has been achieved, collectively these would lead to an overall profile of the HMN (Step 2).

Our own experience to date has been largely confined to these first two steps. However, we have begun to characterize multiple networks as part of Step 3 (see also the Section 5 below), and in so doing, a set of profiles will be created which may be used to compare similar networks. Such cross-network comparisons will potentially identify common features and designs for similar networks (Step 4 in our methodology), revealing parallels in the HCD of HMNs not previously evident.

## 5 Case Trials

The initial typology and framework has been applied to case studies pertaining to ICT innovation and development projects. In the following, we present two of these along with experiences and results. The case trials concern human-centred design processes in which the typology and framework is applied, including a crisis management system and a peer-to-peer reselling network. The cases exemplify how the typology and framework provide increased insight during the HCD phases concerning context analysis, user requirement engineering, and design.

### 5.1 Case 1: Crisis Management System

The eVACUATE project (http://www.evacuate.eu) provides a decision-support system to help operational staff as well as (potentially) emergency services to track and safely guide evacuees in a crisis situation ranging from severe events such as a fire or terrorist threat to less severe operational responses such as responding to overcrowding. Following the initial steps outlined above (**Fig. 3**), we generated two profiles in connection with this case, one representing normal operations when effectively the HMN is simply monitoring activity, and a second for a possible emergency situation. The resulting profiles are shown in **Fig. 4**.

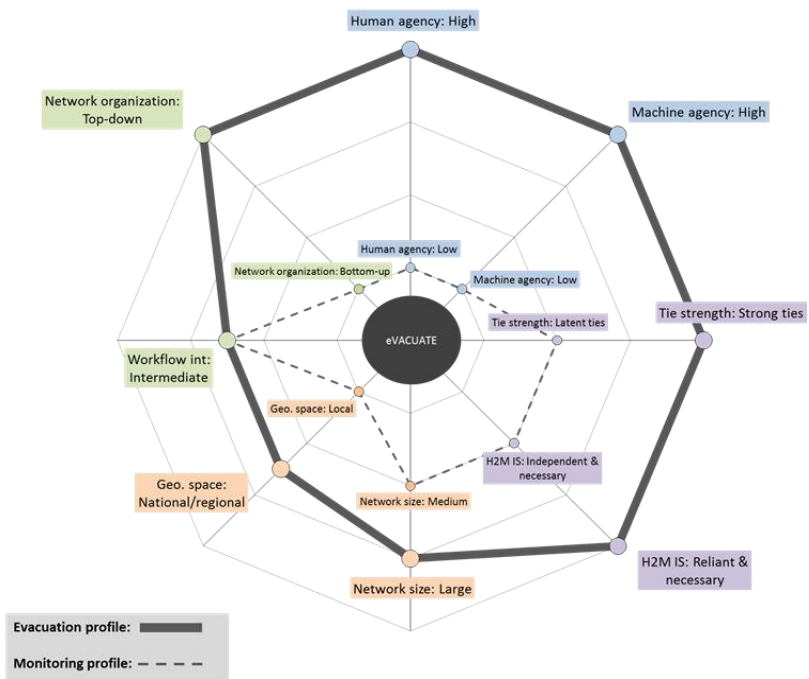

**Fig. 4.** Joint profile for the eVACUATE HMN

For normal operations ("monitoring": the dashed line in the figure), the network is relatively constrained (shown by network size and geographical space), with low human and machine agency: the network as a whole is simply ticking over (low in both cases), checking that the situation is normal and no intervention is required; human interaction with machines is moderate and confined essentially to operational staff (H2M interaction strength: independent and necessary), who need to review machine input (sensor data) and reporting from the decision support system itself. In consequence, tie strength, for instance, remains latent: operational staff has little interaction with members of the public being monitored, who in turn have little interconnection beyond family or friendship groups, and incidental proximity to others.

Contrast this, though, with an emergency situation ("evacuation": the solid line in **Fig. 4**). Focus simply on the machine and human parts of the network: both change dramatically in terms of size, but especially in the level of agency apparent for each. The decision support system may physically increase, for instance, by recruiting additional devices in the crisis situation itself, such as personal communication devices, intelligent signage and so forth, but also by connection from an emergency services network if warranted. Evacuees in extremis will become more interdependent for their own well-being and even survival; operational staff themselves may become evacuees depending on location and situation. In consequence, tie strength becomes 'strong', and human agency 'high'.

An initial validation of the accuracy of the profile was effected through discussion with software engineers with experience of the eVACUATE system and knowledge of the related scenarios. Having introduced them to the rationale behind the profiling framework, they were presented with the profiles in **Fig. 4** and asked for their comments. Although in broad agreement with the dimensions and our interpretation in connection with the eVACUATE system, they believed that other dimensions, such as an indication of machine-to-machine dependence or at least interaction, might be necessary to describe such networks accurately. Notwithstanding such omission and a concern about the scalability of the visualisation itself, the software engineers were positive in two important aspects. First, the profile we provide highlights behavioural aspects of the overall system which may not be apparent from more traditional formal methods: a network, as pointed out by one of the engineers, is much more than its constituent parts. Second, they emphasised that this sort of approach would help communication between the engineers and other 'actors' in the network, namely operational staff and even potential evacuees, in providing a common understanding of the network and what it is there to do. These aspects of the profiling framework are now being investigated further in relation to other HMNs as we build up experience and a repertoire of network types and profiles.

### 5.2 Case 2: Peer-to-Peer Reselling Network

The HMN typology and profiling framework was next applied to the Conserve & Consume project (http://conserveandconsume.wordpress.com/), which focuses on peer-to-peer reselling markets. Here, we have utilized the dimensions of the typology to analyse and discuss a recently launched iOS/Android app for such markets. The app is designed to enable direct selling of goods: sellers snap a photo of the item for sale, add a maximum of 58 character description, and publish the ad. Potential buyers see a thumbnail of the image, the distance to the seller, the price, and the time since the ad was published, and can then choose to open the item to read the description. The analysis has included data from developer and stakeholder collaboration, user interviews, as well as content analyses of classified advertisements posted by users.

The resulting profile is shown in **Fig. 5**. In this paper we will focus on the dimensions network size, geographical space and machine agency.

As is seen in the profiling, network size and geographical space are scored at different values (desired and current profile), reflecting the developer and stakeholder aim of increasing network size and eventually to become an app with global reach.

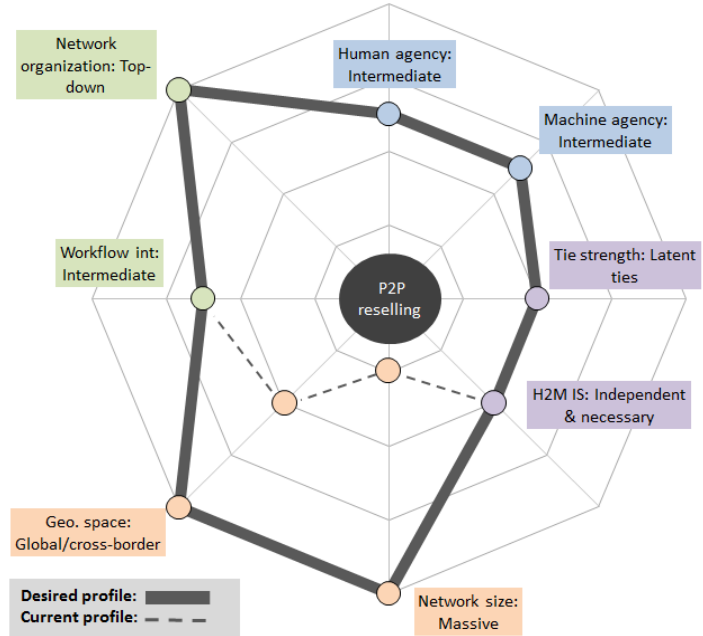

**Fig. 5.** Joint profile for the peer-to-peer reselling HMN[1]

Machine agency is considered to be intermediate, with an important role for human nodes in the network. In its current version, the app performs a small set of functions, notifies users, and influences user-experience. The latter is apparent in how the app is designed to take off as much of the work-load in creating ads as possible, and in the features helping users discover items, and follow peers. End-user interviews demonstrate how users appreciate these functions, making the app functional, efficient and enjoyable both for users as sellers and buyers.

From working with the typology and profiling, several aspects of the network design became apparent as a consequence of how network size, geographical space and machine agency interrelate.

First, the desired increase in network size may require increased, or improved, machine agency to facilitate emergent change. As the group of typical users of the net-work evolves, the categories of items sold through the network also need to evolve. To support this, the developers prioritize functions for automatic classification of content in ads based on image- and text-recognition, and supports automatic updating of filtering categories to keep up with the evolving characteristics of the stuff sold through the marketplace. Hence, the link between machine agency, network size, and network capacity for emergent chance (related to the behaviour layer) has become apparent to the Conserve and Consume project through the profiling activity.

---

[1] The joint profile for Case 2 differs from that presented in the HUMANE technical report *Typology and method v1* [5], due to updates following data collection and collaboration with the users and stakeholders.

Second, whereas the app has a national reach, with some occasional users present also across the world, networks of human nodes seem to gather locally. Transactions between buyers and sellers are made locally, and end-users request improved functions that limit the availability of items for sale to their region. Such user-patterns and preferences call for design supporting the local, even for HMNs that on an aggregate level are global.

Third, the profiling activity provides terminology for discussing characteristics, and enables stakeholders to consider both the current state of an HMN and the desired state. For example, whereas the local will likely remain important, content-analysis of ads indicates heterogeneous user-patterns. Geographical proximity appears important for transactions to be completed, yet analyses of users' ads show that close-by items do not get more views than far-away items.

## 6   Discussion

### 6.1   Lessons Learnt

The application of the proposed typology in the cases described in Section 5 provides useful insights into the potential benefits but also the challenges associated with the dimensions. Significant benefits of the typology include the usefulness of the dimensions to support cross-disciplinary discussion of non-functional aspects of design (Case 1) and the novel understanding of HMN characteristics (Case 2). Furthermore, the profiling procedure was found to support the intended first two steps of defining network characteristics and establishing a network profile effectively.

At the same time, a number of challenges were identified. For instance, developers or stakeholders may desire additional dimensions to accurately describe the HMN. Furthermore, whereas the dimensions were found to support dialogue and aspects of analysis, the reliability in classifying HMNs on the individual dimensions were questioned. Future work is needed both to refine and consolidate the dimensions, and also to improve reliability in analysis for the dimensions.

Another critical challenge derived by the implementation of the dimensions concerned the identification of specific network types based on network profiles, that is, step 3 in the profiling process. This step requires the grouping of similar profiles; the mapping of these profiles would make this easier to carry out. However, the grouping rules and network types' identification remain a challenge. Future work is needed to deploy the typology for matching or clustering HMNs on the basis of their type.

Finally, some early indications have been provided concerning how the profiling framework may support the extraction of design guidelines. For example, in Case 2 considerations regarding machine agency serve to clarify the benefit of increasing automatic support for content classification and categorization; that is, increasing machine agency rather than working towards increased human agency to support growth in HMNs that are comparable to such reselling markets. At the same time, the cases only show early indications as to how such profiling may elicit design knowledge and experience, and future work is needed for the profiling framework to deliver more robust design support for future HMNs.

### 6.2 Limitations

The typology described here is not an exhaustive representation of all possible HMNs. The typology has emerged from and been tested against only a limited number of cases so far, and many more possible configurations will need to be considered as this typology matures. Also, the cases thus far are retrospective analyses of existing HMNs; to be of most value, the typology should also be tested at the design and initiation phases of HMNs to understand the extent to which the typology can help designers achieve their goals for the HMN.

Nevertheless, there are some early signs that there are benefits to this approach based on feedback from people close to the cases where it has been applied. What needs to be done now is work on resolving issues and concerns around both dimensions and methodology to capitalise on such potential. In our initial approach, the risk of overlooking some critical dimensions is inevitable. We attempted to scale each dimension as generally as possible, but the coarseness and the discrete nature of scales in each dimension will limit the applicability of the framework to certain cases.

Even though we believe that our typology can be very useful in understanding HMNs at the theoretical level, we provide no verification of its usefulness as a means to access design support in this paper. This is recognised as a significant gap which needs to be addressed as part of future work.

### 6.3 Conclusion and Future Work

We have presented an initial typology and framework for profiling HMNs in order to support HCD practitioners in the design of more successful systems, focusing on aiding the context analysis and requirements phases. Initial results from case studies in the HUMANE project have demonstrated value, such as helping system designers understand the relationship between aspects of the technical system and its users in order to achieve their vision of their desired system, and improving cross-disciplinary communication when performing requirements elicitation.

In the short term, we will continue to evaluate the typology and profiling framework against more current ICT projects based on superficially different HMNs. The purpose of this is twofold: first, we can thereby extend the investigation of dimensions; and secondly, we will begin to establish a set of profiles for future comparison. On this basis, the typology can then be assessed and validated against a larger set of case studies to ensure it is widely applicable. Network profiling on an even larger scale will pave the way for an analysis into network types and correlations between the dimensions in order to develop an extended technology for identifying a) similar networks and b) relevant design guidance and shared experience. This will address the two final steps of identifying similar networks and extracting design principles in the profiling framework proposed in this paper in order to maximize the potential value of this work to the system designers.

**Acknowledgements.** This work has been conducted as part of the HUMANE project (http://humane2020.eu), which has received funding from the European Union's Horizon 2020 research and innovation programme under grant agreement No 645043.